\documentclass[12pt]{article}
\usepackage{putex}
\usepackage{amsmath}
\usepackage{amssymb}
\usepackage{subfigure}

\usepackage{graphicx}
\usepackage{pstricks}
\usepackage{epsfig}

\def \half {{\frac{1}{2}}}

\begin{document}

\preprint{UCB-PTH-05/23\\
LBNL-58744\\ PUPT-2173\\ hep-th/0508220}

\title{Counter-examples\\[5pt] to the correlated stability conjecture}

\institution{PU}{Joseph Henry Laboratories, Princeton University, Princeton, NJ 08544}

\institution{UCB}{Berkeley Center for Theoretical Physics and Department of Physics, \cr University of California, Berkeley, CA 94720-7300}

\institution{LBL}{Theoretical Physics Group, Lawrence Berkeley National Laboratory, \cr Berkeley, CA 94720-8162}

\authors{Joshua J. Friess,\worksat\PU\ Steven S. Gubser,\worksat\PU\ and Indrajit Mitra\worksat{\UCB,\LBL}}

\abstract{We demonstrate explicit counter-examples to the Correlated
Stability Conjecture (CSC), which claims that the horizon of a black brane is unstable precisely if that horizon has a thermodynamic instability, meaning that its matrix of susceptibilities has a
negative eigenvalue.  These examples involve phase transitions near
the horizon.  Ways to restrict or revise the CSC are suggested.  One of our examples shows that ${\cal N}=1^*$
gauge theory has a second order chiral symmetry breaking phase
transition at a temperature well above the confinement scale.}

\date{August 2005}

\maketitle

\tableofcontents

\section{Introduction}
\label{INTRODUCTION}

In \cite{glOne,glTwo} it was observed that the event horizons
surrounding black strings and $p$-branes are often unstable toward
linear perturbations with some sinusoidal dependence on a spatial
direction parallel to the brane.  According to general theorems,
entropy must increase as such an instability develops.  Indeed,
entropy increase was used in \cite{glOne,glTwo} to motivate the
existence of the instability, but the entropy argument in that case
was that a final state consisting of separated black holes must have
larger entropy than an initial state of a non-extremal uniform
string.  It has subsequently become a difficult question how and
whether one can actually evolve in finite asymptotic time from one
state to the other.  This question was first raised in \cite{hm},
and recent reviews \cite{KolReview, HarmarkOne} provide a summary of what is
presently known as well as a guide to the (already extensive)
literature.

In \cite{gmOne,gmTwo} it was suggested that the existence of a
perturbative Gregory-Laflamme (GL) instability for a horizon which
is infinite and translationally invariant in some spatial direction
should be associated with a local thermodynamic instability.  For
uncharged black strings or branes, this simply means that a GL
instability should occur precisely when the specific heat is
negative.  Analytic arguments in support of this link were advanced
in \cite{reall}.  These arguments are convincing in the case of pure
gravity and for a limited class of charged black branes: see
\cite{RossWiseman} for a recent extension.  A more complicated case
is when the black brane carries some charges or angular momenta
which are capable of being spatially redistributed: see
\cite{gFriess} for a discussion of how Reall's argument might
generalize.  An example is a black string in five or more dimensions
which carries electric charge under a $U(1)$ gauge field $A_\mu$. In
such a case, local thermodynamic stability is a criterion that one
applies to the Hessian matrix of second derivatives of the entropy
with respect to the mass and all the conserved charges.  If this
matrix has a positive eigenvalue, then there is a local
thermodynamic instability.  It should then be possible to locally
redistribute mass density and/or conserved charge density in such a
way as to increase entropy without changing the total value of the
conserved charges.  It makes sense therefore to conjecture that a GL
instability occurs precisely when there is a local thermodynamic
instability.  This is the correlated stability conjecture (CSC) of
\cite{gmOne,gmTwo}.  The conjecture includes as a hypothesis that
the horizon is infinite and translationally invariant in some
spatial direction because only then should we entirely trust
thermodynamic arguments.

Following \cite{reall}, there has been debate over whether the CSC
should be generally valid.  It seems impossible for there to be a
violation in the direction of having a GL instability in the
presence of stable thermodynamics, because then the horizon area
would decrease as one goes forward in time.  Violations in the other
direction do not spoil any broad properties of general relativity,
but (to our knowledge) none have come to light as yet.

In this paper, we find violations of the CSC in the direction where
the arguments of the previous paragraph indicate they are least
likely---but entropy does not decrease!  At the linearized level,
the perturbatively unstable mode involves only a scalar field.  This
scalar acts as an order parameter of a phase transition that occurs
near the black hole horizon.  The violation of the CSC is associated
with trying to keep the black hole in the disordered phase below the
critical temperature at which spontaneous ordering is entropically
favored.  Entropy is expected to increase as the instability departs
from the linear regime, and a definite endpoint of the evolution can
be guessed: it is simply the ordered phase of the black hole
horizon, possibly with some domain walls or other defects within the
brane world-volume.

The instability we find is a normalizable disturbance of the horizon
which grows exponentially in time.  Because it involves only a
scalar field, it may seem rather different from the gravitational
modes investigated in \cite{glOne}.  But there are other examples
\cite{gmOne,gmTwo} where there is a horizon instability involving
only matter fields.  (It can even be shown in these cases that one
can interpolate between an instability involving only the metric and
an instability involving only matter fields.)  The main difference
between the standard Gregory-Laflamme instability and the
instability we study here is the dispersion relation below the
critical wave-number where the frequency $\omega$ becomes imaginary.
In \cite{glOne,glTwo}, the gravitational instabilities were found to
have a characteristic dispersion relation $|\Im \omega| \approx {k_c
\over 10} \sin {\pi k \over k_c}$.  This analytic expression is only
a rough fit to the numerical results of \cite{glOne,glTwo}.  The
feature to note is that $\Im \omega \to 0$ as $k \to 0$.  We may
understand this heuristically as a consequence of conservation of
energy: the $k=0$ mode cannot be excited at all because it would
change the total mass of the black string.  For the instabilities we
will describe, $|\Im \omega|$ is a monotonically decreasing function
of $k$ up to some $k_c$ where it vanishes.  The instability is not
associated with a conserved quantity, so there is no reason for $\Im
\omega$ to vanish as $k \to 0$.

The instability we will describe is somewhat reminiscent of the gyrating strings proposal of \cite{marolf}.  Discussions of this proposed violation of the CSC have focused on non-uniqueness of black brane solutions carrying specified conserved quantum numbers.  This is in the spirit of a near-horizon phase transition
\cite{gPhase}.  What makes a violation of the CSC possible is that knowing the conserved charges of a black brane is not enough to uniquely determine the classical solution.  Local thermodynamic stability has to do only with the conserved quantities, but a dynamical horizon instability is a property of the entire solution.

Although the CSC as stated in \cite{gmOne,gmTwo} seems to us now to be violated by the examples of the present paper, it may be possible to save it by adding an extra hypothesis, namely that the uniform brane solution is unique once all conserved quantities have been specified.

The organization of the rest of this paper is as follows.  In section~\ref{MAGNETIC} we consider a class of examples based on magnetically charged branes with an unusual $\chi^2 F^2$ coupling of a scalar field $\chi$ to a gauge field strength $F$.  In section~\ref{ADS}, we explore an example in $AdS_5$ with two scalars, one of which describes an RG flow, while the other describes the breaking of a spontaneous symmetry.  In section~\ref{DEFORM}, we show that the ${\cal N}=1^*$ deformation of ${\cal N}=4$ SYM exhibits a finite-temperature phase transition, and that a meta-stable branch with no gaugino condensate has a horizon instability in the same class as the examples discussed in section~\ref{ADS}.

\section{Magnetically charged branes}
\label{MAGNETIC}

Consider the action
 \eqn{Maction}{
  S = \int d^D x \, \sqrt{g} \left[
   R - {1 \over 2} (\partial\chi)^2 -
   f(\chi) F_{D-p-2}^2 - V(\chi) \right] \,,
 }
where $V(0)=V'(0)=0$, $f(0)=1$, and $f'(0)=0$.  There is a magnetically charged $p$-brane solution in which $\chi=0$ identically:
 \eqn{pBraneSolutions}{
  ds^2 &= H^{-2/(p+1)} (-h dt^2 + d\vec{x}^2) +
    H^{2/(D-p-3)}
     \left( {dr^2 \over h} + r^2 d\Omega_{D-p-2}^2 \right)
      \cr
  H &= 1 + {r_0^{D-p-3} \sinh^2 \alpha \over r^{D-p-3}} \qquad
  h = 1 - {r_0^{D-p-3} \over r^{D-p-3}}  \cr
  F_{D-p-2} &= \sqrt{2 (D-2) (D-p-3) \over p+1}
    r_0^{D-p-3} \cosh\alpha \sinh\alpha \vol_{S^{D-p-2}} \qquad
 }
where $\vol_{S^{D-p-2}}$ is the volume form on the unit sphere $S^{D-p-2}$.  Special cases of the solutions \pBraneSolutions\ include the magnetically-charged Reissner-Nordstrom black hole ($D=4$, $p=0$), the M2-brane ($D=11$, $p=2$), and the M5-brane ($D=11$, $p=5$).  We will assume $0\leq p \leq D-4$ to avoid certain pathologies, like spacetimes which are not asymptotically flat.

 To check for the sign of specific heat, it is easiest to first
 compute the temperature of the black brane solution in terms of
 $\alpha$ and $r_0$:
 \eqn{temp}{\eqalign{
  T&= {{(D-p-3)} \over {4 \pi r_0}} \left( \cosh \alpha
 \right)^{2\mu} = {{(D-p-3)} \over {\pi r_0 2^{\mu + 2}}} \left( 1+ {\sqrt{1+ {Q^2
 \over r_0^{2(D-p-3)}}}} \right)^{\mu} \,,
}}
where $\mu = -(D-2)/(D-p-3)(p+1)$ and in the last equality we've eliminated the parameter $\alpha$ for
 $Q$ which is proportional to the conserved charge $\int F_{D-p-2}$.
It's straightforward to check that the specific heat, which is inversely
 proportional to and has the same sign as $dT/dr_0$, is positive for these solutions for the
 charge to mass ratio $Q/r_0^{(D-p-3)}$ larger than some $O(1)$ lower bound.  There is no conserved charge that is capable of being spatially redistributed, so positive specific heat means that there is local thermodynamic stability.  In such a situation, the CSC predicts no dynamical instability of the horizon.

Consider however the equation of motion for $\chi$, linearized around \pBraneSolutions:
 \eqn[c]{ChiEOM}{
  \left[ \Box - f''(0) F_{D-p-2}^2 - V''(0) \right] \chi = 0  \cr
  \left[ H^{2/(p+1)} \left( {\omega^2 \over h} - \vec{k}^2 \right) +
   {1 \over r^{D-p-2} H^{2/(D-p-3)}} \partial_r
   \left(r^{D-p-2} h \partial_r \right) -
    m_{\rm eff}^2 \right] \chi_{\omega,\vec{k}} = 0  \cr
  m_{\rm eff}^2 = V''(0) + f''(0)
   {2(D-2)(D-p-3) \over p+1} r_0^{2(D-p-3)}
    {H^{-2 (D-p-2) / (D-p-3)} \over r^{2(D-p-2)}}
    \cosh^2 \alpha \sinh^2 \alpha \,,
 }
where in the second line we have specialized to a separated $s$-wave ansatz: $\chi = e^{-i\omega t + i\vec{k} \cdot \vec{x}} \chi_{\omega,\vec{k}}(r)$.  If $V''(0) < 0$, then empty space is unstable toward development of a VEV for $\chi$.  This instability is also visible in the stress tensor as a violation of the dominant energy condition.  Let us therefore assume $V''(0) \geq 0$.  If we choose $f''(0)$ sufficiently negative, then \ChiEOM\ admits normalizable solutions with imaginary $\omega$.  This is a GL instability.  Provided $f(\chi) > 0$ for all $\chi$, the dominant energy condition is not violated.  Thus our main conclusion: violations of the CSC can be arranged by choosing suitable couplings of a scalar field.

To verify the claims in the previous paragraph about the dominant energy condition, let us examine the stress tensor explicitly:
 \eqn[c]{FullT}{
  T_{\mu\nu} = T_{\mu\nu}^{(\chi)} + T_{\mu\nu}^{(F)}  \cr
  T_{\mu\nu}^{(\chi)} = \partial_\mu \chi \partial_\nu \chi - 
    {1 \over 2} g_{\mu\nu} (\partial\chi)^2 - g_{\mu\nu} V(\chi)  \cr
  T_{\mu\nu}^{(F)} = f(\chi) \left[ {1 \over (D-p-3)!}
    F_{\mu\mu_2 \cdots \mu_{D-p-2}} 
    F_\nu{}^{\mu_2 \cdots \mu_{D-p-2}} - 
     {1 \over 2 (D-p-2)!} g_{\mu\nu} F_{\mu_1 \cdots \mu_{D-p-2}}^2
       \right]
 }
The dominant energy condition is that when $\xi^\mu$ is timelike, $T_{\mu\nu} \xi^\mu \xi^\nu \geq 0$ (heuristically: energy density is positive) and $T^\mu{}_\nu \xi^\nu$ is timelike or null (heuristically: the flow of energy is timelike or null).  It is well known that $T_{\mu\nu}^{(\chi)}$ obeys dominant energy provided $V(\chi) \geq 0$.  Assume also that $f(\chi) \geq 0$.  Then $T^{(F)}_{\mu\nu}$ is a non-negative multiple of its form in the special case where $f(\chi) = 1$.  This special case is a free massless theory with a positive definite hamiltonian density, and it is intuitively clear that the flow of energy is timelike or null for any such theory.  A formal demonstration that $T^{(F)\mu}{}_\nu \xi^\nu$ is timelike or null for timelike $\xi^\mu$ is slightly technical and will not be presented here. 

What drives the existence of a normalizable eigenfunction
$\chi_{\omega,\vec{k}}$ with negative $\omega^2$ is that $m_{\rm
eff}^2$ becomes large and negative near the horizon when $f''(0)$ is
sufficiently negative, holding all other quantities (including
$\vec{k}^2$) fixed.  Heuristically $\chi$ becomes tachyonic near the
horizon, so it should condense.  To verify this more rigorously, we
may transform \ChiEOM\ into Schr\"odinger form.  We assume $r_0>0$.
The extremal case is somewhat different on account of the nature of
horizon boundary conditions, but the same qualitative claim about
condensation of $\chi$ in the presence of sufficiently negative
$f''(0)$ should persist at extremality.

The radial equation in \ChiEOM\ can be cast in the form
 \eqn{ChiBetter}{
  \left[ \omega^2 + {J_1 h \over r^{D-p-2}} \partial_r
    \left( r^{D-p-2} h \partial_r \right)
     - h \left[ J_2 V''(0) + \vec{k}^2 + K_1 f''(0) \right]
       \right] \chi_{\omega, {\vec k}} = 0 \,,
 }
where the functions $J_i$ are explicitly given by:
 \eqn{Jdef}{
 J_1 &= H^{{-2(D-2) \over (D-p-3)(p+1)}}  \qquad
 J_2 = H^{{-2 \over p+1}} \cr
 K_1 = {2(D-2)(D-p-3) \over p+1} &r_0^{2(D-p-3)}
    {H^{2(p^2+4 p-d (p+2)+5) / (D-p-3)(p+1)} \over r^{2(D-p-2)}}
    \cosh^2 \alpha \sinh^2 \alpha
}
The functions $J_i$, considered on the range $r_0 \leq r < \infty$, are analytic, bounded above and below by positive numbers, with a limit of $1$ as $r \to \infty$.  The function $K_1$, considered on the same range, is analytic, everywhere positive, bounded above, and with a finite non-zero limit at the horizon.  Evidently, either class of functions may be added, multiplied, and raised to real powers without going outside that class of functions.  Define
 \eqn{NewDefs}{
  u &= \int^r {d\tilde{r} \over P} \qquad
   P = h \sqrt{J_1}  \cr
  \tilde\chi &= {\chi_{\omega,\vec{k}} \over F} \qquad
   F = r^{-{D-p-2 \over 2}} \sqrt[4]{J_1} \,,
 }
where the lower limit on the integration defining $u$ is set so that $u-r \to 0$ as $r\to \infty$.  Note that $u \to -\infty$ at the horizon, because $h$ has a simple zero there as a function of $r$.  The radial equation becomes
 \eqn[c]{ChiBest}{
  \left[ -\partial_u^2 + V(u) \right] \tilde\chi =
    \omega^2 \tilde\chi  \qquad
  V = h \left[ J_2 V''(0) + \vec{k}^2 + K_1 f''(0) \right] +
   (\partial_u \log F)^2 -
    \partial_u^2 \log F \,.
 }
The extra terms $(\partial_u \log F)^2 - \partial_u^2 \log F$ introduced into $V$ by the transformation of variables are uniformly bounded.  They tend to $0$ exponentially fast as $u\to -\infty$, and so does $V(u)$ itself.  And they tend to $0$ as $u\to \infty$ as well, but $V$ has a positive non-zero limit there, namely $V''(0) + \vec{k}^2$.  Moreover, $V(u)$ is analytic, and each of its derivatives is uniformly bounded over the real line.

Most importantly, by making $f''(0)$ sufficiently negative, $V(u)$
can be made as negative as one pleases over any specified finite
range of $u$, without altering its smoothness or asymptotic
properties at $\pm\infty$.  It is then a standard fact of quantum
mechanics that an $L^2$-normalizable bound state appears in the
spectrum for sufficiently negative $f''(0)$.  Indeed, any number of
such bound states, each corresponding to a different negative value
of $\omega^2$, can be introduced into the spectrum by sending
$f''(0)$ more and more negative.  The eigenfunction for a negative
$\omega^2$ bound state decays exponentially as $u\to \pm\infty$. The
existence of such a bound state to the Schr\"odinger problem
\ChiBest\ implies the existence of an
exponentially growing perturbation of $\chi$ which is finite outside the horizon, regular at the horizon, and exponentially small at infinity: a GL instability.

Using \ChiBest\ one may also argue in general that if all parameters
except $\vec{k}^2$ are held fixed, and $\vec{k}^2$ is increased
sufficiently, there is no GL instability.  It is generally
understood that when a GL instability exists for small $\vec{k}^2$, it persists up to a critical value $k_c^2$, and at $k_c^2$ there is a static, normalizable perturbation of the horizon.  Demonstrating this from
\ChiBest\ involves a minor technical complication: for $\omega=0$, there isn't an $L^2$-normalizable eigenfunction $\tilde\chi(u)$, but instead an eigenfunction which approaches a constant as $u \to -\infty$, is uniformly bounded, and decays exponentially as $u \to \infty$.  The corresponding $\chi_{0,\vec{k}}$ is finite at the horizon (so clearly regular there), finite outside the horizon, and exponentially small at infinity---hence normalizable.\footnote{We have not given a precise definition of normalizability of a perturbation in the scalar field $\chi$, but several could be used interchangeably in the current context.  Most physically, the norm could be taken as the energy density in the perturbation integrated over a slice of constant~$t$ outside the horizon.  Alternatively, $\chi_{\omega,\vec{k}}(r)$, considered as a function on a slice of constant~$t$ outside the horizon, could be required to be in all $L^p$ spaces for $p>1$.  The ``minor technical complication'' amounts to showing that a function $\tilde\chi(u)$ which is not $L^2$-normalizable nevertheless translates into a perturbation $\chi_{0,\vec{k}}(r)$ which is normalizable in either of the senses just described.

An additional feature generally required of a perturbation in order for it to be a physically meaningful classical instability is that there should be no outgoing energy flux at the horizon.  This is true of solutions with negative $\omega^2$ because $\chi_{\omega,k}(r) \to 0$ at the horizon, and it is true of the $\omega=0$ solution because $\dot\chi = 0$.}

Unlike other examples of GL instabilities, in this case it is fairly
clear that there are possible static endpoints of the evolution with
uniformly controlled curvatures.  In particular, there are static,
spatially uniform solutions with $\chi$ non-zero but vanishing
asymptotically as $r\to \infty$.  This non-zero profile of $\chi$ is
an example of hair for the black brane, because it is not determined
by any conserved quantity.  Static hairy solutions were studied in
quantitative detail for the case $p=0$, $D=4$ in \cite{gPhase}, and
the qualitative features should be the same in other cases.  In
particular, it is not necessary for $V(\chi)$ or $f(\chi)$ to have
extrema at non-zero $\chi$ in order for static hairy solutions to
exist.  If $V''(0) > 0$, then for given magnetic charge, hairy
solutions exist only for $r_0$ sufficiently small.  The hair
develops smoothly as $r_0$ crosses a critical value $r_c$: slightly
below this critical value, $\chi(r)$ in the hairy solution is
uniformly small outside the horizon.  The hairy solutions, when they
exist, have greater entropy than the solutions \pBraneSolutions.  It
is in this respect that the current example of a GL instability is
especially reminiscent of gyrating branes \cite{marolf}: the driving
intuition in that case was that for a certain range of parameters,
there was a more entropically favorable way to carry certain quantum
numbers than the standard stationary black string solution.

If there is a ${\bf Z}_2$ symmetry of the classical action under
$\chi \to -\chi$, then it should also be possible in principle to
construct static domain wall solutions where a black brane has non-zero $\chi$ which is negative on one side of the domain wall
(say, as $x^1 \to -\infty$) and positive on the other (say, as $x^1
\to \infty$).  In fact, a variety of more or less intricate solitons
can be contrived, depending on the topological structure of the space of solutions with spatially uniform horizons and a given set of conserved charges.

It is clear that the GL instability and hairy black brane solutions under discussion relate to second order phase transitions on the world-volume of the brane.  The dictionary between the gravitational description and the world-volume theory is clearest when the spacetime is asymptotically anti-de Sitter rather than asymptotically flat.  We therefore turn to this case in the next section.

\section{An example in $AdS_5$}
\label{ADS}

The horizon instability discussed in the previous section is driven
by the $\chi^2 F^2$ coupling.  This sort of coupling can arise in
low-energy effective actions of compactified string theory, but it
is not particularly familiar to us from other contexts.  It is
therefore interesting to look for violations of the CSC in systems
that are well studied for other reasons.  A class of examples in
$AdS_5$ was suggested in \cite{gPhase}.  The action is
\eqn{AdSaction}{\eqalign{
 S &= \int d^5 x {\sqrt g} \left[R - {1 \over 2} (\partial \phi)^2 - {1
 \over 2} (\partial \chi)^2 - V(\phi,\chi) \right] \, \cr
 V(\phi,\chi) &= -{12 \over L^2} + {1 \over 2} m_{\phi}^2 \phi^2 +{1
 \over 2} m_{\chi}^2 \chi^2 + g \phi^2 \chi^2 \,, \quad \quad g<0 \,.
 }}
This action is still ``made up,'' but in section~\ref{DEFORM} we will see that an interesting two-dimensional slice of the scalar manifold of $d=5$, ${\cal N}=8$ gauged supergravity gives an action of essentially this form.

We are interested in black brane solutions in the Poincar\'e slice of $AdS_5$.  The metric is
 \eqn{metric}{
 ds^2 = e^{2A(r)}\left[-h(r)dt^2 + d{\vec x}^2 \right] + {{dr^2} \over
 h(r)} \,.
 }
It is assumed that $\phi$ has a non-zero profile corresponding to a deformation of the CFT lagrangian.   $\chi$, on the other hand, may be zero or non-zero, but its asymptotics near the boundary of $AdS_5$ are required to indicate a VEV of the dual operator in the CFT rather than a deformation by it.  It was demonstrated in \cite{gPhase} for a particular choice of $m_\phi$, $m_\chi$, and $g$ that there is a continuous transition from a ``disordered phase'' where $\chi=0$ (so that $\langle {\cal O}_\chi \rangle = 0$) to an ``ordered'' phase where $\chi\neq 0$ (so that $\langle {\cal O}_\chi \rangle \neq 0$).  If we denote the value of $\phi$ at the horizon by $\phi_0$, then the transition happens for a particular value $\phi_c$ of $\phi_0$: for $\phi_0 < \phi_c$ one is in the disordered phase, and for $\phi_0 > \phi_c$ one is in the ordered phase.

The action \AdSaction\ has a ${\bf Z}_2 \times {\bf Z}_2$ symmetry, associated with $\phi \to -\phi$ and $\chi \to -\chi$.  The first ${\bf Z}_2$ is broken explicitly by the boundary conditions on $\phi$ near the boundary of $AdS_5$.  The second ${\bf Z}_2$ is preserved in the ordered phase and broken in the disordered phase.

The lagrangian of the dual field theory is
 \eqn{DualFT}{
  {\cal L} = {\cal L}_{\rm CFT} +
   \Lambda_\phi^{4-\Delta_\phi} {\cal O}_\phi
 }
where $\Lambda_\phi$ is an energy scale and $\Delta_\phi = 2 +
\sqrt{4 + m_\phi^2 L^2}$ is the dimension of ${\cal O}_\phi$
(assumed to be less than $4$).  ${\cal L}_{\rm CFT}$ is the
lagrangian of the undeformed conformal field theory.  In
section~\ref{DEFORM} we will encounter a specific example where the
CFT is ${\cal N}=4$ super-Yang-Mills theory and ${\cal O}_\phi$ is a
fermion mass term---up to certain subtleties to be mentioned below.
The Hawking temperature of the black hole horizon translates into a
finite temperature for the field theory.  It may be assumed that
 \eqn[c]{Asymptotic}{
  A(r) \to {r \over l} \qquad h(r) \to 1  \cr
  \phi(r) \to X_1 e^{(\Delta_\phi - 4) r/L} +
    X_2 e^{-\Delta_\phi r/L} \qquad
  \chi(r) \to Y_2 e^{-\Delta_\chi r/L}
 }
as $r \to \infty$.\footnote{The asymptotics for $\phi(r)$ can be more complex, because the $e^{(\Delta_\phi - 4) r/L}$ behavior may be modified by subleading exponentials which are nevertheless larger than $e^{-\Delta_\phi r/L}$.  This will not be a concern for present purposes.}  $X_1$ is proportional to $\Lambda_\phi^{4-\Delta_\phi}$, and it may be set equal to unity by making an appropriate dilation of $\vec{x}$ and $t$.

Black brane solutions of the form \metric\ with the asymptotics \Asymptotic\ may be parametrized by $\phi_0$, or alternatively in dual field theory language, by $T/\Lambda_\phi$.  The precise relation between these two functions can be established through numerics.  In particular, the critical value $\phi_c$ is associated with a critical temperature $T_c$ at which the phase transition occurs.

Our goals in the remainder of this section are
 \begin{enumerate}
  \item to demonstrate that there is a Gregory-Laflamme instability for black branes with $\phi_0 > \phi_c$;
  \item to extract two critical exponents of the phase transition.
 \end{enumerate}
In addressing point~1 we will be satisfied to show that, for black brane backgrounds with $\phi_0 > \phi_c$ and $\chi=0$, there is a normalizable, static mode in the linearized equation of motion for $\chi$ with a non-zero wave-number $k_c$.  To understand point~2 we will solve the linearized equation for $\chi$ and extract $\langle {\cal O}_\chi(x) {\cal O}(0) \rangle$.  The full equation of motion for $\chi$ is
 \eqn{perturb}{
  \Box \chi = {{\partial V} \over \partial \chi} \,,
 }
and with a separated ansatz $\chi = e^{i \vec{k} \cdot \vec{x}} \tilde\chi(r)$ one obtains
\eqn{kpert}{
 e^{-4 A(r)} \left[-k^2 e^{2 A(r)} + {d \over {dr}} \left(h(r) e^{4 A(r)}
 {d \over dr}\right) \right]{\tilde \chi}(r) = \left[m_\chi^2 L^2 + 2 g
 \phi^2(r) \right]{\tilde \chi}(r) \,.
}
Near the horizon ($r=0$), one of the solutions to \kpert\ diverges
 logarithmically with $r$, while the finite solution asymptotes to a
 constant:
\eqn{kHorizon}{
 {\tilde \chi}(r) \to {\tilde \chi}_h \left(1 + {k^2 + m_\chi^2 L^2 +
 2 g \phi_0^2 \over h'(0)} r \right) \,.
 }
Far away from the black hole, ${\tilde \chi}(r)$ dies off as
 \eqn{kFalloff}{
 {\tilde \chi}(r) \to Y_1(k) e^{(\Delta_\chi - 4)r/L} + Y_2(k)
 e^{-\Delta_\chi r/L} \,.
 }
Because the equation for $\tilde\chi(r)$ is linear, $Y_1(k)$ and $Y_2(k)$ are not individually meaningful.  But their ratio,
 \eqn{GkDef}{
  G(k) \equiv Y_2(k) / Y_1(k) \,,
 }
is well-defined, and up to a $k$-independent constant, it is the scaling part of the correlator $\langle {\cal O}_\chi (\vec{k}){\cal O}_\chi (-\vec{k}) \rangle$ evaluated in the thermal state.  (This correlator also may include analytic terms in $k$, corresponding to contact terms.  These are excluded by construction in $G(k)$.)

\begin{figure}[h]
  \hfill
  \begin{minipage}[t]{.45\textwidth}
    \begin{center}
      \epsfig{file=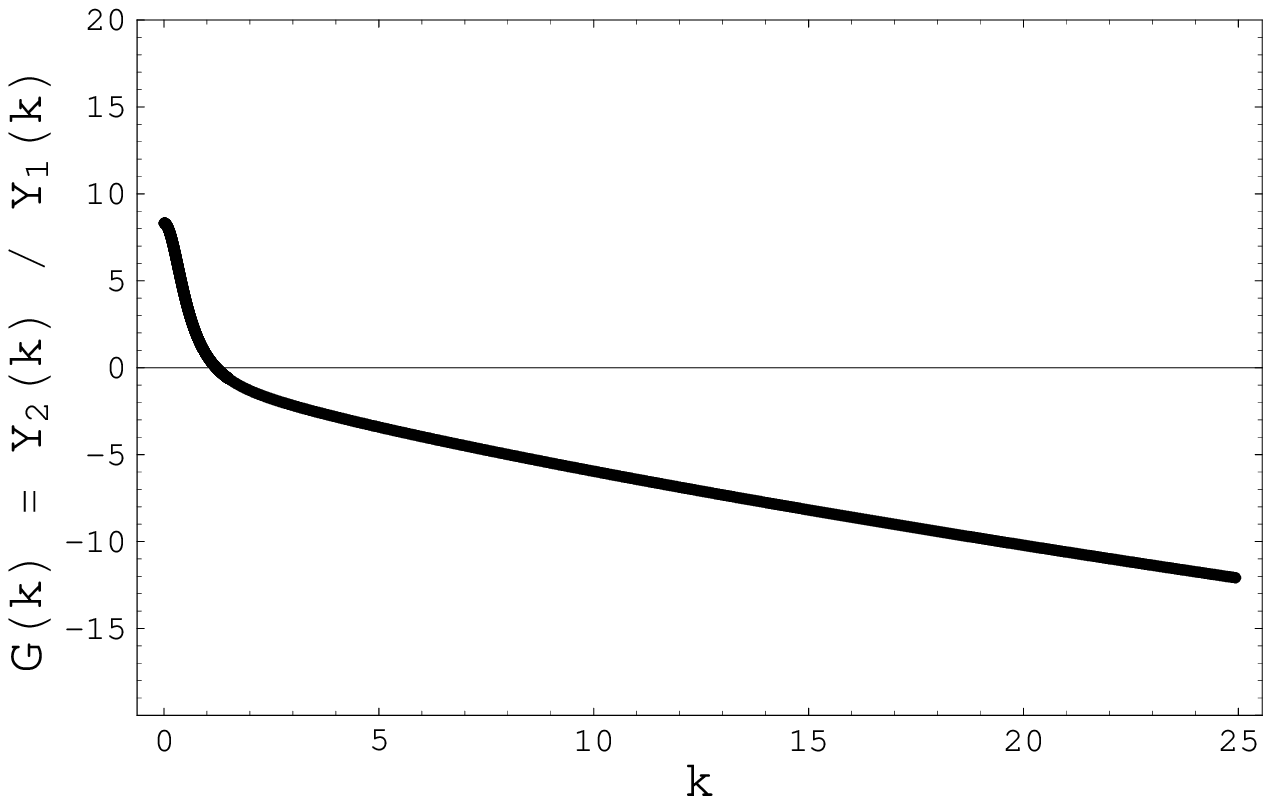, scale=0.5}
      \caption{Propagator for $\phi_0$~$=$~$0.95$~$\phi_c$, where
        $\phi_c = 0.428$.  The lack of a singularity
        indicates the absence of a normalizable mode.}
      \label{FigA}
    \end{center}
  \end{minipage}
  \hfill
  \begin{minipage}[t]{.45\textwidth}
    \begin{center}
      \epsfig{file=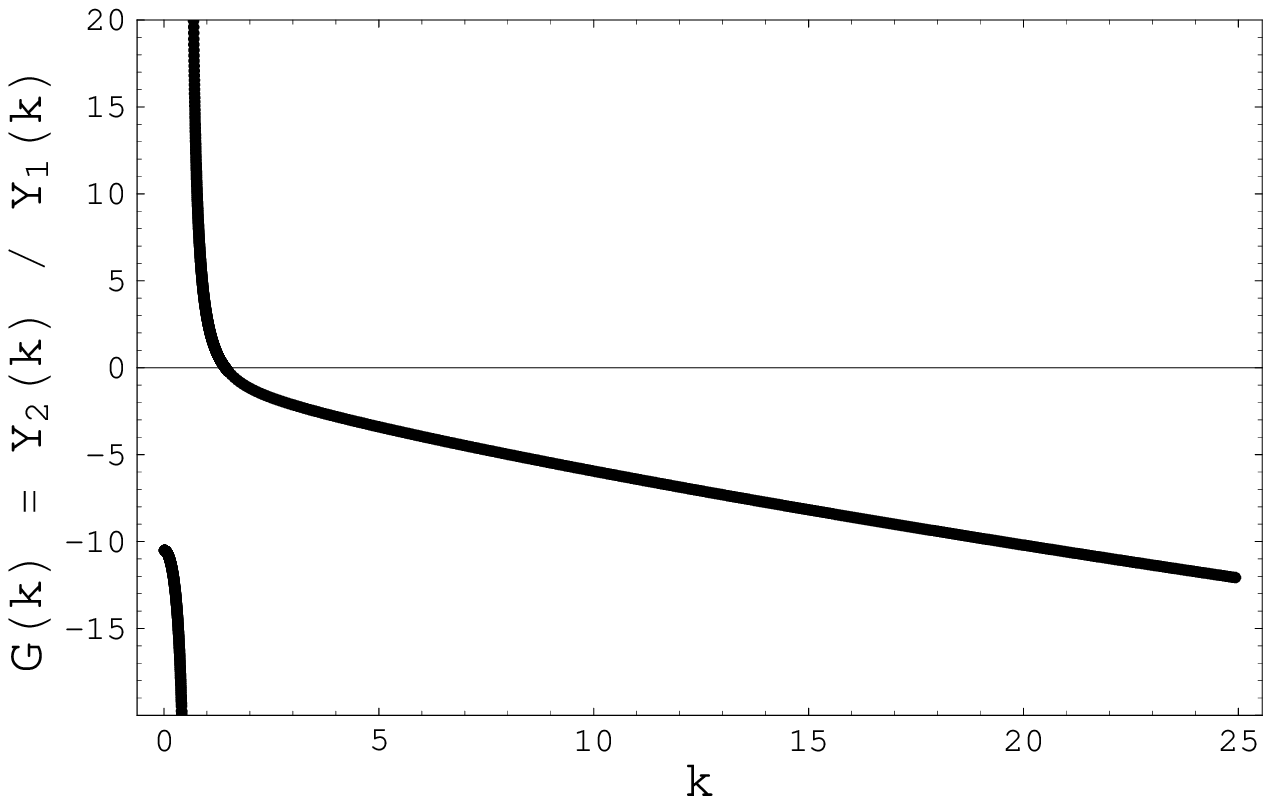, scale=0.5}
      \caption{Propagator for $\phi_0=1.05 \phi_c$.  The singularity indicates a normalizable, stationary mode, and the existence of a GL instability for $k < k_c$.}
      \label{FigB}
    \end{center}
  \end{minipage}
  \hfill
\end{figure}

Figures~\ref{FigA} and~\ref{FigB} show $G(k)$ for
 \eqn{StandardChoice}{
  g=-10 \quad \Delta_\phi = 2.2 \quad \Delta_\chi = 2.4 \,,
 }
on either side of the phase transition.  On the disordered side, the singularity at a finite wave-number $k_c$ signals the existence of a GL instability.  As $k \to k_c$, $Y_1(k) \to 0$ while $Y_2(k)$ remains finite.  So at $k=k_c$, the static mode $\chi = e^{i \vec{k} \cdot \vec{x}} \tilde\chi(r)$ is normalizable.

\begin{figure}[h]
  \centerline{
   \epsfig{file=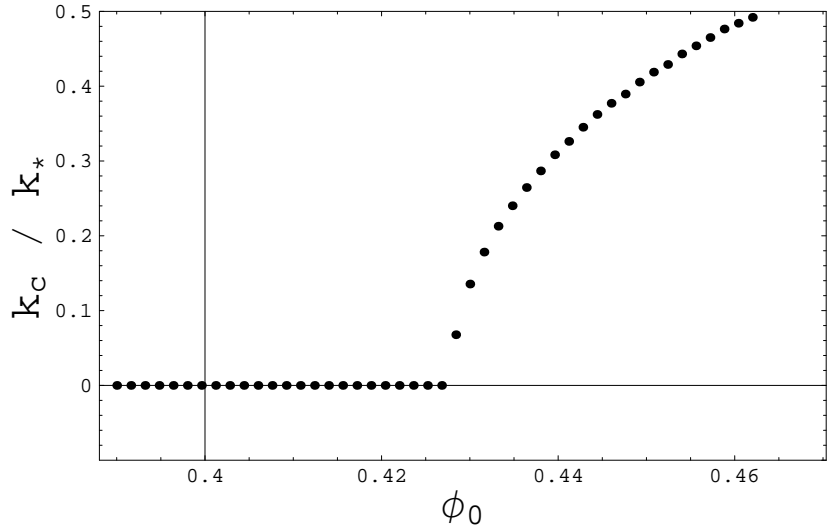, scale=0.8}
  }
  \caption{Plot showing non-zero critical values of $k_c$ for $\phi_0 > \phi_c$ (the ordered phase).  $k_*$ is the value of $k$ at which $G(k) = 0$.}
  \label{FigC}
 \end{figure}

As one approaches the phase transition, the static mode's wave-number $k_c$ should go to zero: this is the standard way in which a GL instability appears or disappears \cite{gOzakin}.  In Figure~\ref{FigC} we show that indeed this happens for the choice $g = -10$, $\Delta_\phi = 2.2$, and $\Delta_\chi = 2.4$.  It is convenient to plot the ratio $k_c/k_*$, where $k_*$ is the wave-number at which the scaling part of $G(k)$ vanishes.  As $\phi \to \phi_c$ from above, $k_*$ has a finite limit, while
 \eqn{kkcExponent}{
  {k_c \over k_*} \sim \left( {T_c - T \over T_c} \right)^\zeta
   \qquad \zeta \approx 0.479 \,.
 }
Given the accuracy of the fits, this result for the exponent $\zeta$ is consistent with $\zeta = 1/2$, which was also found (approximately) for the GL instability of non-extremal D3-branes, M2-branes, and M5-branes in \cite{gOzakin}.

Another critical exponent can be defined as follows: let $\overline{\cal O}_\chi(x) = T \int_0^\beta dt_E \, {\cal O}_\chi(t_E,x)$.  Then for $x \gg 1/T$, one expects a power-law behavior
 \eqn{ObarPower}{
  \langle \overline{\cal O}_\chi(x)
    \overline{\cal O}_\chi(0) \rangle \sim
    {1 \over x^{2\overline\Delta_\chi}} \,.
 }
The expectation value in \ObarPower\ is taken in the thermal state right at the critical point ($\phi_0=\phi_c$), and the dimension $\overline\Delta_\chi$ is distinct from the dimension $\Delta_\chi$.  Note that the correlator in \ObarPower\ is the Fourier transform in only the three spatial directions $\vec{x}$ of $\langle {\cal O}_\chi(\vec{k}) {\cal O}_\chi(-\vec{k}) \rangle$: we work with zero Matsubara frequency throughout.  Carrying out the Fourier transform, one obtains
 \eqn{OkPower}{
  \langle {\cal O}_\chi(\vec{k}) {\cal O}_\chi(-\vec{k}) \rangle
    \sim k^{2\overline\Delta_\chi - 3} \,,
 }
again for $T=T_c$.

\begin{figure}[h]
  \centerline{
   \epsfig{file=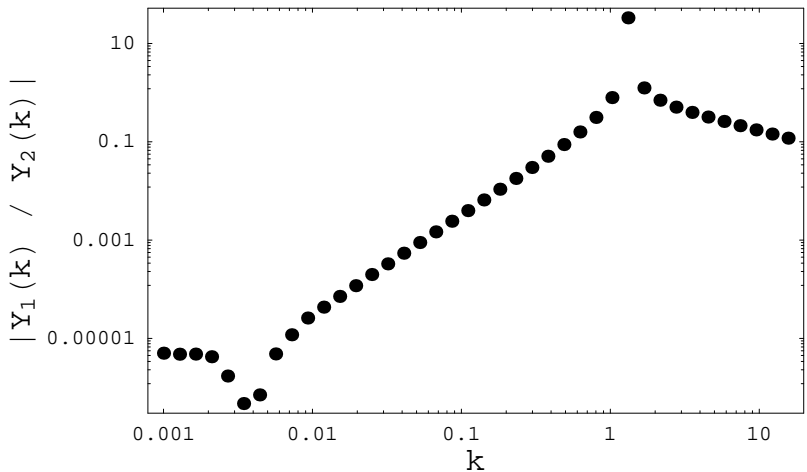, scale=0.8}
  }
  \caption{Inverse propagator for $(\phi_0-\phi_c)/\phi_0 \approx 10^{-6}$.}\label{FigD}
 \end{figure}

In testing the power-law prediction \OkPower\ and obtaining the critical exponent, it is convenient for numerics to depart slightly from $T=T_c$ and examine a scaling region of $k$.  In Figure~\ref{FigD} we show an example with $\phi_0$ slightly greater than $\phi_c$.  The scaling region is cut off in the infrared by $k_c$, below which there is a GL instability.  And it is cut off in the ultraviolet by $k_*$, above which there is a gradual transition to a different power law, controlled by the dimension $\Delta_\chi$ pertaining to the vacuum state.  The numerical results for the slope are consistent with $\overline\Delta_\chi = 1/2$.

A notable feature of the present discussion is that we see scaling behavior for $x \gg 1/T$ without manifest conformal invariance in a similar limit.  Conformal invariance in the supergravity background would be associated with recovering a four-dimensional anti-de Sitter space by making a Kaluza-Klein reduction in the Euclidean time direction---a procedure that amounts to restricting attention to zero Matsubara frequency.  But such a reduction leads instead to a singular background where the scalar that controls the size of the circle in the Euclidean time direction diverges.

\section{A finite temperature phase transition in ${\cal N}=1^*$ super-Yang-Mills theory}
\label{DEFORM}

So far, all of the examples of systems which violate the CSC have relied
on lagrangians without a string-theoretic origin.  In this section
we will explore an example drawn from type~IIB string theory on
$AdS_5 \times S^5$---or, more precisely, from its consistent
truncation to $d=5$, ${\cal N}=8$ gauged supergravity. The dual
field theory is the so-called ${\cal N}=1^*$ deformation of ${\cal
N}=4$ super-Yang-Mills theory.  The relevant deformation gives equal
masses to three of the four adjoint fermion fields.  The scalar
$\phi$ dual to this dimension~$3$ deformation is non-zero because of
the asymptotic boundary conditions at the boundary of $AdS_5$. There
is another $SU(3)$-invariant, dimension $3$ operator: it includes a
bilinear of the fourth adjoint fermion, which is the ${\cal N}=1$
partner of the gauge boson.  The scalar $\chi$ dual to this operator
has asymptotic boundary conditions that allow it to be zero or
non-zero.  If it is non-zero, the VEV of the dual operator ${\cal
O}_\chi$ can be read off from it.  This VEV breaks a chiral
symmetry.  The relevant part of the $d=5$, ${\cal N}=8$ supergravity
lagrangian is \cite{GPPZthree}
  \eqn{StringEx}{
   S &= \int d^5 x \sqrt{g}\left[R - \half (\partial \phi)^2 -
   \half(\partial \chi)^2 - V(\phi, \chi)\right] \cr
   V(\phi, \chi) &=
   -\frac{3}{2}\left(\cosh^2\frac{\phi}{\sqrt{3}} +
   4\cosh\frac{\phi}{\sqrt{3}}\cosh \chi - \cosh^2
   \chi + 4\right) \,.
  }

In short, the situation is very much as in the examples of section~\ref{ADS}.  And the outcome is similar, too: if the horizon is sufficiently close to the boundary of $AdS_5$---corresponding to a small value of $\phi_0$ at the horizon---then $\chi$ and hence $\langle {\cal O}_\chi \rangle$ must vanish.  But if $\phi_0 > \phi_c \approx 4.41$, the background with zero $\chi$ has a Gregory-Laflamme instability.  There is a second-order transition at $\phi_0 = \phi_c$ to backgrounds with $\chi \neq 0$.  The two-point function of ${\cal O}_\chi$ shows the same $1/k^2$ behavior in an infrared scaling region.  These conclusions still hold if we replace the transcendental function $V(\phi,\chi)$ by a polynomial approximation to it of the form considered in section~\ref{ADS}:
 \eqn{PolynomialV}{
   L^2 V = -12 - \frac{3}{2}\left(\phi^2 +
 \chi^2\right) - \frac{\phi^2 \chi^2}{2} \,.
 }
In the language of section~\ref{ADS}, $g = -1/2$.  With the modified potential \PolynomialV, the critical horizon value of $\phi_0$ is $\phi_c \approx 2.96$.

\begin{figure}
  \centerline{
   \epsfig{file=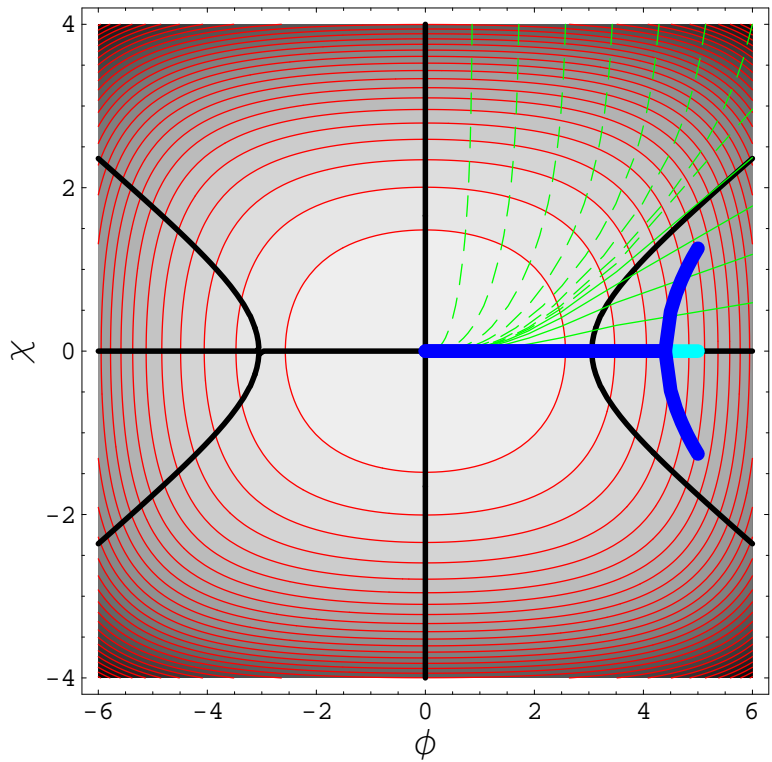, scale=1.0}
  }
  \caption{The thick blue lines show the values $(\phi_0,\chi_0)$ of the scalars at the horizons we were able to produce numerically.  Dark blue indicates stable solutions.  Light blue indicates solutions with a GL instability.
  The thin green trajectories are the holographic RG flows \SeveralFlows.  They are found most simply as the gradient flows of a superpotential $W$, whose contours are shown in red.  The aspect ratio of this figure is not 1:1, so it is not readily apparent that the green trajectories are orthogonal to the red contours.
The $C=0$ trajectory is asymptotic to the black curve, which is part of the locus where the gradient of $W$ is parallel to the gradient of $V$.  The $C<0$ trajectories are shown in solid green, and the $C>0$ trajectories are shown in dashed green.}
  \label{FigE}
 \end{figure}

In \cite{GPPZthree}, holographic RG flows were considered of the form
 \eqn[c]{SeveralFlows}{
  ds^2 = e^{2A(r)} (-dt^2 + d\vec{x}^2) + dr^2 \qquad
   \phi = \phi(r) \quad \chi = \chi(r)  \cr
  \phi = \sqrt{3} \log {1 + e^{-r} \over 1 - e^{-r}}
   \qquad
  \chi = \log {1 + e^{-3r + C} \over 1 - e^{-3r + C}} \,.
 }
These geometries are singular in the infrared for all values of $C$.
It is hard to know which of them is physical.  In \cite{gNaked}, it
was suggested that the ones with $C > 0$ are unphysical, that the
$C=0$ trajectory is probably physical, and that the $C < 0$
trajectories might also be physical.  The reason to think this is
that the $C > 0$ trajectories cannot be limits of backgrounds with
regular horizons, while the $C=0$ solution is in a special class
that was conjectured in \cite{gNaked} to be precisely the limits of
backgrounds with regular horizons.  Our numerical results are
consistent with these conjectures: for given $\phi_0 > \phi_c$, only
one background with positive $\chi_0$ was found, and the points
$(\phi_0,\chi_0)$ are not far from the $C=0$ trajectory---see
Figure~\ref{FigE}.  Only a limited range of $\phi_0$ was explored,
so we cannot be sure that the $C=0$ trajectory is the limit of black
hole solutions.  For larger $\phi_0$, numerical noise prevented
reliable results.

It may be that the solutions with regular horizons that we have found are the only stable ones in the temperature range under consideration, even if we go beyond $d=5$, ${\cal N}=8$ gauged supergravity.  Other solutions would correspond to turning on scalars in $AdS_5$ which have positive $m^2$.  Standard no-hair arguments suggest that this is impossible: only one of the two solutions to the linearized equations for such scalars is normalizable at infinity, and horizon boundary conditions generically prevent this normalizable mode from being turned on.  Also, the total scalar potential must be more negative than $-12/L^2$ at the horizon \cite{gNaked}, which suggests that large deformation by positive mass scalars is impossible.

In sum, the second order chiral symmetry breaking phase transition exhibited in this section gives a tantalizing first glimpse of the finite-temperature behavior of ${\cal N}=1^*$ super-Yang-Mills theory.  Much more is known about its zero-temperature phases \cite{VafaWitten,DonagiWitten} and their ten-dimensional description in terms of five-branes \cite{PolchinskiStrassler}.  The fact that the chiral symmetry breaking transition happens when there is a regular horizon with an entropy scaling as $N^2$ shows that this transition is well above the confinement scale.  This is somewhat reminiscent of the model of \cite{Manohar:1983md} where unconfined quarks interact with a chiral condensate.

\section{Conclusions}
\label{CONCLUSIONS}

Although the correlated stability conjecture (CSC) of
\cite{gmOne,gmTwo} has served as a useful guide to the
Gregory-Laflamme (GL) instability in various settings, we have
argued that it fails to correctly predict horizon instabilities that
are unrelated to conserved quantities.  Instead, these instabilities
are associated (in the examples we have described) with second order
phase transitions in which the unstable horizon represents the
disordered phase cooled below the critical temperature at which
ordering should take place.  The ordered phase is represented by a
new (and presumably stable) uniform black brane solution.

The nature of these counter-examples leads us to conjecture that the
CSC works provided that there is a unique background with a
spatially uniform horizon and specified conserved charges.  We know
of no counter-examples to this restricted version of the conjecture.

A simple refinement of the CSC has been suggested to us by B.~Kol:\footnote{We thank B.~Kol for his permission to explain this proposal here.} thermodynamic stability should be redefined by enlarging the Hessian matrix to include derivatives of the entropy with respect to quantities that characterize the asymptotics of scalar fields.  (In the examples given, the asymptotics of $\chi$ are characterized by a single real parameter: $Y_2$ in the example of section~\ref{ADS}, if one sets $X_1=1$: see \Asymptotic.)  Local thermodynamic stability then amounts to the absence of a positive eigenvalue for this enlarged Hessian matrix, and the revised version of the CSC is that GL instabilities occur precisely when local thermodynamic stability is lost, provided the brane has infinite volume and translation invariance in some spatial direction.  It seems very likely that this refinement of the CSC survives all tests to date.

While we have not shown explicitly that there is a violation of the
CSC (as originally phrased in \cite{gmOne,gmTwo}) in an asymptotically flat background of a well-defined string
theory, the there should be no difficulty in principle in extending
the example of section~\ref{DEFORM} to an asymptotically flat background: one simply has to ``re-attach'' flat space to the
asymptotically $AdS_5 \times S^5$ throat region.  The resulting
background and its unstable perturbation should still possess a
global $SU(3)$ symmetry, which may make it easier to find them
explicitly.

For second order phase transitions in $AdS_5$, we found in an
infrared scaling region, where spatial separations are much greater
than $1/T$, that the dimension of the operator ${\cal O}_\chi$
(whose VEV is the order parameter) is $1/2$.  Over the rather wide
set of parameter choices that we checked, this result is independent
of the dimension of ${\cal O}_\chi$ at the ultraviolet fixed point,
as well as the dimension of ${\cal O}_\phi$ and the coupling
constant $g$. We believe this result can be understood as a
consequence of large $N$: when normalized to have an $O(1)$
two-point function, the higher point functions of ${\cal O}_\chi$
are suppressed by factors of $N$.\fixit{Better be sure}  So the
state created by ${\cal O}_\chi$ in the three-dimensional effective
theory describing physics in the infrared scaling region is, up to
$1/N$ corrections, a free scalar, and the dimension of a free scalar
in three dimensions is $1/2$.

\section*{Acknowledgements}

The research of J.~F.~was supported in part by the NSF Graduate Research Fellowship Program, and by the Department of Energy under Grant No.\ DE-FG02-91ER40671.  J.~F.\ would like to thank H.~Verlinde for a useful discussion and the organizers of TASI~2005 for their
hospitality while this work was in progress.

S.~G.\ would like to thank I.~Klebanov for a useful discussion and B.~Kol for commenting on an early version of the paper.  The research of S.~G.\ was supported in
part by the Department of Energy under Grant No.\ DE-FG02-91ER40671, and by the Sloan Foundation.

I.~M. was supported in part by the Director, Office of Science, Office of High Energy and Nuclear Physics, of the U.S. Department of Energy under Contract~DE-AC02-05CH11231, in part by the National Science Foundation under grant PHY-00-98840, and by the Berkeley Center for Theoretical Physics. I.~M.\ would also like to thank the Aspen Center for Physics for hospitality during the final stages of this work.

\bibliographystyle{ssg}
\bibliography{GLhair}

\end{document}